\documentclass[twoside]{article}
\usepackage{graphicx}
\usepackage{fancyhdr}
\usepackage{multicol}
\usepackage{styl-ap}

\newcommand{\la}{\raisebox{-0.6ex}{$\,\stackrel
{\raisebox{-.2ex}{$\textstyle <$}}{\sim}\,$}}
\newcommand{\ga}{\raisebox{-0.6ex}{$\,\stackrel
{\raisebox{-.2ex}{$\textstyle >$}}{\sim}\,$}}

\begin{document}
\title{The space density of magnetic and non-magnetic cataclysmic variables, and implications for CV evolution}
\author{M.L. Pretorius}
\workplace{Department of Physics, University of Oxford, Denys Wilkinson Building, Keble Road, Oxford OX1 3RH, United Kingdom
\next
Previous address: School of Physics and Astronomy, University of Southampton, Highfield, Southampton SO17 1BJ
}
\mainauthor{retha.pretorius@astro.ox.ac.uk}
\maketitle

\begin{abstract}%
We present constraints on the space densities of both non-magnetic and
magnetic cataclysmic variables, and discuss some implications for
models of the evolution of CVs.  The high predicted non-magnetic CV
space density is only consistent with observations if the majority of
these systems are extremely faint in X-rays.  The data are consistent
with the very simple model where long-period IPs evolve into polars
and account for the whole short-period polar population. The fraction
of WDs that are strongly magnetic is not significantly higher for CV
primaries than for isolated WDs. Finally, the space density of IPs is
sufficiently high to explain the bright, hard X-ray Galactic Centre
source population.
\end{abstract}


\begin{multicols}{2}
\section{Introduction}
There are still many uncertainties in the theory of cataclysmic
variable (CV) formation and evolution, as well as several serious
discrepancies between predictions and the properties of the observed
CV population (e.g.\ Patterson 1998; Pretorius, Knigge \& Kolb 2007a;
Pretorius \& Knigge 2008a,b; Knigge, Baraffe \& Patterson 2011). In
order to constrain evolution models, more and better observational
constraints on the properties of the Galactic CV populations are
needed. A fundamental parameter predicted by evolution theory, that is
expected to be more easily measured than most properties of the
intrinsic CV population, is the space density, $\rho$. A few specific,
important open questions concerning the formation and evolution of
CVs:
\begin{itemize} 
\item[(i)] Is the large predicted population of non-magnetic CVs at
  short orbital period consistent with the current observed CV sample?
\item[(ii)] Is there an evolutionary relationship between IPs and
polars?
\item[(iii)] Can the intrinsic fraction of mCVs be reconciled with the incidence of
magnetic WDs in the isolated WD population? 
\item[(iv)] Do mCVs dominate the total Galactic X-ray source
populations above $L_X \sim 10^{31}\,{\rm erg s^{-1}}$?
\end{itemize}
These questions can be addressed empirically, with reliable
measurements of the space densities of the different populations of
CVs.

\begin{figure}[tbh]
\centerline{\resizebox{120mm}{!}{\includegraphics{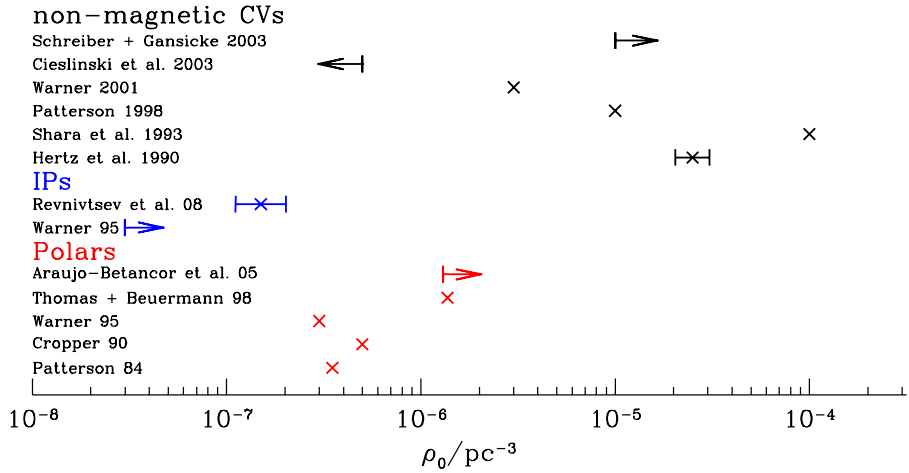}}}
\caption{Some previously reported measurements of the space densities of different CV populations.}
\label{fig:previous}
\end{figure}

Uncertainty in $\rho$ measurements is in part caused by
statistical errors, arising from uncertain distances and small number
statistics. However, the dominant source of uncertainty is most likely
systematic errors caused by selection effects. Fig.~\ref{fig:previous} shows some reported measurements (differing by several orders of magnitude for non-magnetic CVs). 

Selection effects are most easily accounted for in samples with
simple, well-defined selection criteria.  In the absence of a useful
volume-limited CV sample, a purely flux-limited sample is the most
suitable for measuring $\rho$. Whereas optical CV samples always
include selection criteria based on, e.g., colour or variability,
there are X-ray selected CV samples that are purely flux-limited. All
active CVs show X-ray emission generated in the accretion
flow. Furthermore, mCVs are luminous X-ray sources, while the
correlation between the ratio of optical to X-ray flux and the optical
luminosity of non-magnetic CVs, implies that an X-ray flux limit does
not introduce as strong a bias against short-period CVs as an optical
flux limit (e.g.\ van Teeseling et al.\ 1996).

Here we use 2 X-ray surveys, the \emph{ROSAT} Bright Survey (RBS;
e.g.\ Schwope et al.\ 2002), and the \emph{ROSAT} North Ecliptic Pole
(NEP) survey (e.g. Henry et al.\ 2006) to construct X-ray flux-limited
CV samples. We then provide robust observational constraints on the
space densities of both magnetic and non-magnetic CVs, by carefully
considering the uncertainties involved.

We provide additional background on the questions listed above in
Section~\ref{sec:context}, present the measurements in
Section~\ref{sec:results}, discuss the implications of the results in
Section~\ref{sec:discussion}, and finally list the conclusions in
Section~\ref{sec:concl}.

\end{multicols}
\begin{multicols}{2}

\section{Context}
\label{sec:context}

\subsection{Missing non-magnetic CVs}
It is not clear whether the present-day observed non-magnetic CV population is inconsistent with theoretical expectations. Population synthesis models predict that only $\simeq 1$ percent of all CVs are above the period gap (see e.g.\ Kolb 1993). The the vast majority of CVs should therefore be intrinsically faint. Pretorius et al. (2007a) and Pretorius \& Knigge (2008b) used a specific model of Kolb (1993), together with models of the outburst properties and SEDs of CVs, to show that, although observed CV samples are strongly biased against short-period systems, an as yet undetected faint CV population cannot dominate the overall population to the extent predicted by this particular population synthesis model. Knigge et al.\ (2011) used the properties of CV donor stars to conclude that the AML rate is lower above the gap and higher below the gap than predicted by the standard model. This leads to larger predicted factions of both period bouncers and long-period systems. Whether this is consistent with observed CV samples is not yet known.

That a large faint population of CVs exists is now clear from observations (Gansicke et al.\ 2009; Patterson 2011). However, whether observations have truly revealed a population as large as predicted remains to be seen.

Some predicted values of the non-magnetic CV space density are as high as $2 \times 10^{-4}\,\mathrm{pc^{-3}}$ (de Kool 1992; Kolb 1993); observational estimates are typically much lower, but have a large range; values from $\le 5 \times 10^{-7}\mathrm{pc}^{-3}$ to $\rho \sim 10^{-4}\mathrm{pc}^{-3}$ have been reported. Perhaps the most straight forward test of these predictions is to compare them to the measured space density of the Galactic short-period, non-magnetic CV population. 

\subsection{Evolutionary relationship between IPs and polars}

\end{multicols}
\begin{figure}[tbh]
\centerline{\resizebox{120mm}{!}{\includegraphics{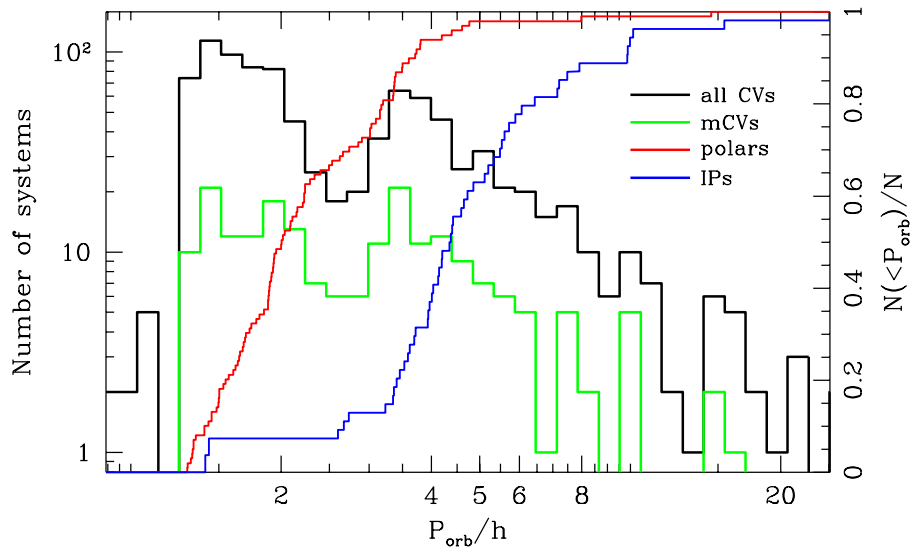}}}
\caption{The orbital period distribution of all CVs (black), and mCVs (green). Cumulative distributions are also shown for polars (red) and IPs (blue). Almost all IPs are found at long $P_{orb}$, while most polars are short-$P_{orb}$ systems. Periods from Ritter \& Kolb (2003).}
\label{fig:perhis}
\end{figure}
\begin{multicols}{2}

In many ways, the formation and evolution of magnetic and non-magnetic
CVs is thought to be similar. Both types of systems form through
common envelope (CE) evolution, evolve first from long to short
$P_{orb}$ because of angular momentum loss (AML), and eventually
experience period bounce, when the thermal time-scale of the donor
becomes longer than its mass loss time-scale. In fact, the main
proposed difference between the evolution of mCVs and non-magnetic CVs
affects only the polars, where magnetic braking (MB) is likely to be
suppressed (e.g.\ Li \& Wickramasinghe 1998; Townsley \& Gansicke
2009). The $P_{orb}$ distributions of magnetic and non-magnetic CVs
are broadly in line with these ideas: if polars and IPs are considered
jointly, their $P_{orb}$ distribution is very similar to that of
non-magnetic CVs, showing a period gap in the range $2\,{\rm hr}
\la P_{orb} \la 3\,{\rm hr}$ and a period minimum at around $P_{orb}
\simeq 80\,{\rm min}$ (see Fig.~\ref{fig:perhis}).

It has been known for a long time that most IPs are found above the
period gap and most polars below (Fig.~\ref{fig:perhis})., which
immediately suggests that IPs may evolve into polars (e.g.\ Chanmugam
\& Ray 1984).  This is a physically appealing idea, since smaller
orbital separation and lower $\dot{M}$ (besides large magnetic field
strength) favour synchronization. MB drives much higher mass-transfer
rates above the period gap than gravitational radiation (GR) does
below; therefore, it is plausible that many accreting magnetic WDs may
only achieve synchronization once they have crossed the period gap.

The main problem with this scenario is that the magnetic fields of the
WDs in IPs ($B_{IP} \la 10 {\rm MG}$) are systematically weaker than
those of the WDs in polars ($B_{polar} \sim 10 - 100 {\rm MG}$). There
are several possible resolutions to this problem. Perhaps the simplest
(in terms of binary evolution) is that the high accretion rates in IPs
could partially ``bury'' the WD magnetic fields, so that the
observationally inferred field strengths for these systems are
systematically biased low (Cumming 2002). It is also possible that the
short-period polar population is dominated by systems born below the
period gap (this still requires an explanation for the fate of
long-period IPs, although they may simply become unobservable; see
Patterson 1994; Wickramasinghe, Wu \& Ferrario 1991).

One way to shed light on the relationship between IPs and polars is
via their respective space densities. For example, if all long-period
IPs evolve into short-period polars, and all short-period polars are
the progeny of long-period IPs, then their space densities should be
proportional to the evolutionary time-scale associated with these
phases. In this particular example, we would predict that
$\rho_{polar}/\rho_{IP} \simeq \tau_{GR}/\tau_{MB} >> 1$.

\subsection{Intrinsic fraction of mCVs}
Magnetic systems make up $\simeq 20\%$ of the known CV population
(Ritter \& Kolb 2003).  At first sight, this is a surprisingly high fraction,
given that the strong magnetic fields characteristic of IPs and polars
($B \ga 10^6\,{\rm G}$) are found in only $\simeq 10\%$ of
isolated WDs (e.g.\ Kulebi et al.\ 2009).  If these numbers are
representative of the intrinsic incidence of magnetism amongst CVs and
single WDs, the difference between them would have significant
implications: either strong magnetic fields would have to favour the
production of CVs, or some aspect(s) of pre-CV evolution would have to
favour the production of strong magnetic fields (see e.g.\ Tout et al.\ 2008).

However, it is actually by no means clear yet that magnetism is really
more common in CV primaries than in isolated WDs. The main problem is
that the observed fraction of magnetic systems amongst known CVs is
almost certainly affected by serious selection biases. For example,
since mCVs are known to be relatively X-ray bright, they are likely to
be over-represented in X-ray-selected samples. Conversely, polars, in
particular, are relatively faint in optical light (since they do not
contain optically bright accretion disks), so they are likely to be
under-represented in optically-selected samples. Given that the
overall CV sample is a highly heterogeneous mixture of X-ray-,
optical- and variability-selected sub-samples (which also usually lack
clear flux limits), it is very difficult to know how the observed
fraction of mCVs relates to the intrinsic fraction of magnetic WDs in
CVs.

\subsection{Galactic X-ray Source Populations}
There have been many attempts to determine the make-up and luminosity
function of Galactic X-ray source populations in a variety of
environments, including the Milky Way as a whole, the Galactic Centre, the Galactic
Ridge, and globular clusters. Remarkably, in all of these environments, mCVs have
  been proposed as the dominant population of X-ray sources above $L_X
  \ga 10^{31}{\rm erg s^{-1}}$.

In most of these studies, the breakdown of the observed X-ray source
samples into distinct populations is subject to considerable
uncertainty, since few of the sources have optical counterparts and/or
properties that permit a clear classification. Identifications of
observed sources with physical populations rely mainly on X-ray
luminosities and hardness, and statistical comparisons of observed and
expected number counts. The local space densities of the relevant
populations are arguably the most important ingredient in these
comparisons. In effect, the question being asked is whether the
extrapolation of the local space density to the environment being
investigated can account for the observed number of sources seen
there. In the case of mCVs, such extrapolations are difficult,
primarily because the local space densities are rather poorly
constrained.

\section{Calculating space densities}
\label{sec:results}

\subsection{The flux-limited samples}
The RBS covers $|b|>30^\circ$ to $F_X \ga
10^{-12}\mathrm{erg\,cm^{-2}s^{-1}}$, and includes 16 non-magnetic
CVs, and 30 mCVs (6 IPs and 24 polars). The NEP covers 81~sq.deg. to
$F_X \ga 10^{-14}$. Only 4 CVs where detected in the NEP, all of them
non-magnetic. The samples are presented in Pretorius et al.\ (2007b),
Pretorius \& Knigge (2012), and Pretorius et al.\ (2013).


\subsection{The method}
We use the $1/V_{max}$ method (e.g. Stobie et al. 1989) together with
a Monte Carlo simulation designed to sample the full parameter space
allowed by the data, as described in Pretorius et al.\ (2007b) and
Pretorius \& Knigge (2012). We tested the method to verify that it
gives reliable error estimates, and also considered various possible
systematic biases (Pretorius \& Knigge 2012; Pretorius et al.\ 2013).

\subsection{Results}

\subsubsection{Probability distribution functions}

\begin{figure}[tbh]
\centerline{\resizebox{140mm}{!}{\includegraphics{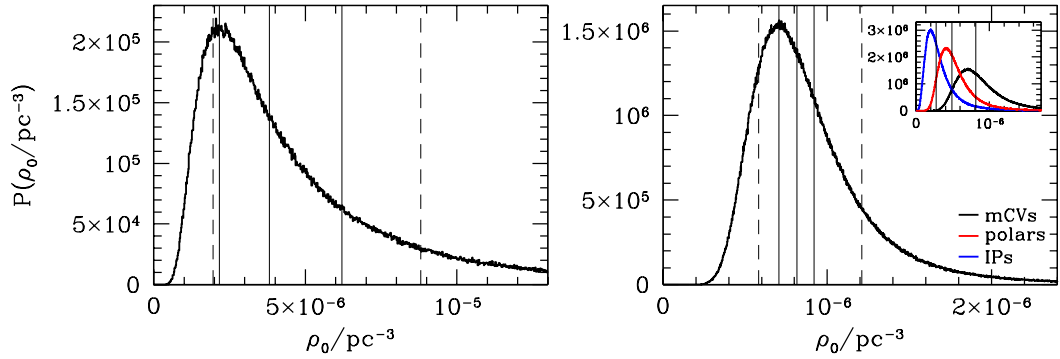}}}
\caption{The $\rho$ distributions for non-magnetic CVs (left-hand panel) and mCVs (right-hand panel) resulting from our simulations.  Solid lines in both panels mark the modes, medians, and means of the distributions; dashed lines show 1-$\sigma$ intervals. The probability distribution functions shown in the inset are for the whole mCV sample, polars alone (red), and IPs alone (blue). Reproduced from Pretorius \& Knigge (2012) and Pretorius, Knigge \& Schwope (2013).}
\label{fig:rhopdf}
\end{figure}

The distributions of mid-plane $\rho$ values, normalized to give probability
distribution functions, from the simulations are shown in
Fig.~\ref{fig:rhopdf}. The best-estimate mid-plane space densities are
$4^{+6}_{-2} \times 10^{-6}\,\mathrm{pc^{-3}}$ for non-magnetic CVs,
and $8^{+4}_{-2} \times 10^{-7}\,\mathrm{pc^{-3}}$ for mCVs. For the 2
classes of mCVs, we find $3^{+2}_{-1} \times
10^{-7}\,\mathrm{pc^{-3}}$ for IPs and $5^{+3}_{-2} \times
10^{-7}\,\mathrm{pc^{-3}}$ for polars.

\subsubsection{Upper limits on $\rho$ of undetected populations}
\label{sec:limits}
The $\rho$ estimates assume that the detected populations are
representative of the underlying population, in the sense that they
contain at least 1 of the faintest systems present in the
intrinsic populations. It is possible that even large populations
of sources at the faint ends of the luminosity functions can go
completely undetected in flux-limited surveys.  

\begin{figure}[tbh]
\centerline{\resizebox{140mm}{!}{\includegraphics{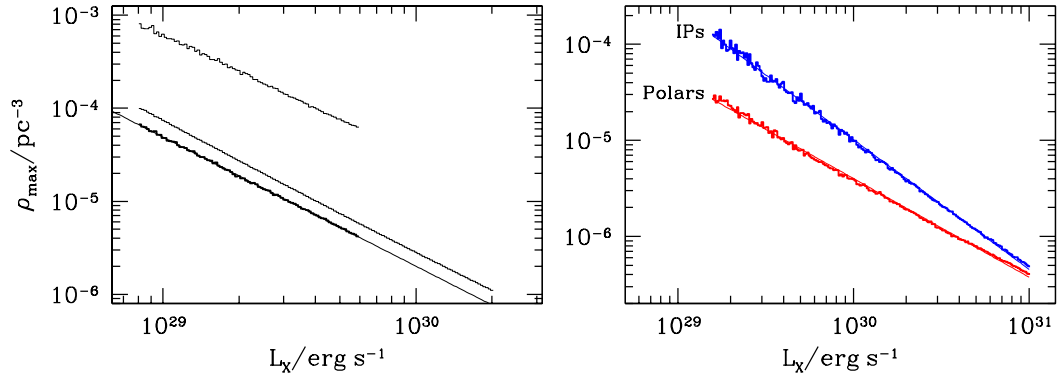}}}
\caption{The upper limit on $\rho$ as a function of X-ray luminosity for undetected populations of CVs.  The left-hand panel is for non-magnetic CVs, and the right-hand panel for IPs (blue) and polars (red). In the left-hand panel, the 2 upper, fine histograms show the corresponding results for the RBS (middle) and NEP (top) surveys alone. Note that the assumed X-ray spectra of non-magnetic CVs, polars, and IPs are different, hence the different slopes. Reproduced from Pretorius \& Knigge (2012) and Pretorius et al.\ (2013).}
\label{fig:limit}
\end{figure}

We performed additional Monte Carlo simulations to place limits on the
sizes of faint populations of CVs that could escape detection in the
surveys we have used (see Pretorius \& Knigge 2012; Pretorius et
al.\ 2013). Fig.~\ref{fig:limit} shows the maximum allowed $\rho$ as a
function of $L_X$, separately for possible undetected non-magnetic,
polar and IP populations.  Specifically, if $\rho_{n-m} = 2 \times
10^{-4}\,\mathrm{pc^{-3}}$ (at the high end of the predicted range),
we require that the majority of non-magnetic CVs have $L_X \la 4
\times 10^{28}\,\mathrm{erg\,s^{-1}}$. A population of undetected
polars with a space density as high as $5 \times$ the measured
$\rho_{polar}$ must have $L_X \la 10^{30}\,\mathrm{erg s^{-1}}$. A
hidden population of IPs can only have $\rho = 5 \times \rho_{IP}$ if
it consists of systems with X-ray luminosities fainter than $5 \times
10^{30}\,\mathrm{erg s^{-1}}$.

\end{multicols}
\begin{multicols}{2}

\end{multicols}
\begin{multicols}{2}

\section{Discussion}
\label{sec:discussion}

\subsection{Missing non-magnetic CVs}
We discuss our measured $\rho_{n-m}$, as well as the upper limit, in the context of the predicted (i) large total space density of non-magnetic CVs, and (ii) large predicted fraction of normal short-period CVs and period bouncers.

Population synthesis models predict that at most a few percent of all CVs are above the period gap (Kolb 1993 finds less than 1\%, while Knigge et al.\ 2011 predict 3\%). Although we find that long-period systems account for slightly more than 50\% of our total $\rho_{n-m}$, the data do not rule out these theoretical predictions. For example, using the Knigge et al.\ (2011) fraction of long-period systems, and assuming that we have not significantly under-estimated the space density of long-period CVs, the space density of short-period CVs is $\simeq 2 \times 10^{-6}\,\mathrm{pc^{-3}}(97/3) \simeq 6 \times 10^{-5}\,\mathrm{pc^{-3}}$. Using the upper limit on $\rho_{n-m}$ from Section~\ref{sec:limits}, we find that a short-period CV population of this size could escape detection in the two surveys, provided that the systems have $L_X \la 8 \times 10^{28}\,\mathrm{erg\,s^{-1}}$ (for the simple case of a hypothetical single-$L_X$ population of faint, undetected CVs). To reach the predicted $\rho = 2 \times 10^{-4}\,\mathrm{pc^{-3}}$ would require that the majority of CVs has X-ray luminosities below $L_X = 4 \times 10^{28}\,\mathrm{erg\,s^{-1}}$.

\subsection{Evolutionary relationship between IPs and polars}
If one assumes that long-period IPs are the sole progenitors of
short-period polars, and that all IPs synchronize once they have
crossed the period gap, then the ratio of the space densities of
long-$P_{orb}$ IPs and short-$P_{orb}$ polars ($\rho_{IP,lp}$ and
$\rho_{polar,sp}$) should simply reflect their relative evolutionary
time-scales. The observed logarithm of this ratio is
$\log{(\rho_{polar,sp}/\rho_{IP,lp})} = 0.32 \pm 0.36$.  If the
evolution of long-period IPs is driven by MB, while that of
short-period polars is driven solely by GR, the evolutionary
time-scale of short-period polars should be $\ga 5 \times$ that of
long-period IPs (e.g. Knigge et al.\ 2011).  This is completely
consistent with the ratio of the inferred space densities. In fact, at
2-$\sigma$, the uncertainties are large enough to encompass both
ratios exceeding 10 and ratios less than unity. This means that, with
the currently available space density estimates for polars and IPs, we
cannot place strong constraints on the evolutionary relationship
between the two classes. Nevertheless, it is interesting to note that
the simplest possible model, in which short-period polars derive from
long-period IPs, is not ruled out by their observed space densities.

\subsection{Intrinsic fraction of mCVs}
Combining the space density estimates of magnetic and non-magnetic CVs
to estimate the intrinsic fraction of mCVs, we find
$\log(f_{mCV}) = -0.80^{+0.27}_{-0.36}$.  This is consistent, within
the considerable uncertainties, with the fraction of isolated WDs that
are strongly magnetic .  In fact, it seems likely that the
X-ray-selected CV sample is more complete for mCVs than it is for
non-magnetic CVs. Therefore, the incidence of magnetism is not
obviously enhanced amongst CV primaries compared to isolated WDs.

\subsection{Galactic X-ray Source Populations}
We consider if it is plausible that IPs dominate X-ray source
populations above $L_X \simeq 10^{31} {\rm erg s^{-1}}$, taking the
Galactic Centre as an example.  The deep Chandra survey of Muno et
al.\ (2009) includes $\simeq 9000$ sources down to $L_X \simeq
10^{31}\,{\rm erg s^{-1}}$, in an area of $\simeq 10^{-3} {\rm
  deg^{2}}$.  Approximating the volume covered by the survey as a
sphere of radius $150$~pc, the space density of X-ray sources in the
Galactic Centre is $\rho_{X,GC} \sim 6\times 10^{-4}\,{\rm pc^{-3}}$,
while the local space density of IPs is $\rho_{IP} \sim 3\times
10^{-7}\,{\rm pc^{-3}}$. However, the stellar space density in the
Galactic centre is $\simeq 1600 \times$ higher than in the solar
neighborhood. Thus these densities are consistent, and we conclude
that IPs remain a viable explanation for most of the X-ray sources
seen in the Galactic Centre.

\section{Conclusions}
\label{sec:concl}
With the assumption that the CV samples from the RBS and NEP surveys
are representative of the intrinsic populations (in the sense that we
detected at least one system at the faintest ends of the luminosity
functions of those populations), we find $\rho_{n-m}=4^{+6}_{-2}
\times 10^{-6}\,\mathrm{pc^{-3}}$ and $\rho_{mCV}= 8^{+4}_{-2} \times
10^{-7}\,\mathrm{pc^{-3}}$ ($\rho_{polar}=5^{+3}_{-2} \times
10^{-7}\,\mathrm{pc^{-3}}$ and $\rho_{IP}=3^{+2}_{-1} \times
10^{-7}\,\mathrm{pc^{-3}}$).

The data are consistent with more than half of non-magnetic CVs having
$28.7<\mathrm{log}(L_X/\mathrm{erg\,s^{-1}})<29.7$, and escaping
detection. However, to reach $\rho_{n-m} = 2 \times
10^{-4}\,\mathrm{pc^{-3}}$ (at the high end of the predicted range),
we require that the majority of non-magnetic CVs have $L_X \la 4
\times 10^{28}\,\mathrm{erg\,s^{-1}}$. 

The ratio of the space density of short-period polars to long-period
IPs is consistent with the very simple hypothesis that long-period IPs
evolve into short-period polars, and that this accounts for the whole
population of short-period polars.

Existing data cannot rule out that strongly magnetic WDs have the same
incidence amongst CVs as in the field.

When the local space density of IPs is scaled to the density of stars
in the Galactic Centre, it is sufficiently high to account for the
number of bright X-ray sources detected in that region.

\thanks 
I thank the organizes for a successful meeting, and for inviting me to present this review.

\end{multicols}

\begin{thebibliography}{99}
\bibitem{} Chanmugam G., Ray A.: 1984, ApJ, 285, 252 
\bibitem{} Cumming A.: 2002, MNRAS, 333, 589
\bibitem{} de Kool M.: 1992, A\&A, 261, 188 
\bibitem{} G{\"a}nsicke B.~T., et al.: 2009, MNRAS, 397, 2170 
\bibitem{} Henry J.P., et al.: 2006, ApJS, 162, 304 
\bibitem{}Knigge C.: 2006, MNRAS, 373, 484 
\bibitem{}Knigge C., Baraffe I., Patterson J.: 2011, ApJS, 194, 28
\bibitem{}Kolb U.: 1993, A\&A, 271, 149
\bibitem{} K{\"u}lebi B., et al.: 2009, A\&A, 506, 1341 
\bibitem{}Li J., Wickramasinghe D.~T.: 1998, MNRAS, 300, 718 
\bibitem{} Muno M.~P., et al.: 2009, ApJS, 181, 110 
\bibitem{} Patterson J.:  1984, ApJS, 54, 443
\bibitem{} Patterson J.: 1994, PASP, 106, 209 
\bibitem{} Patterson J.: 1998, PASP, 110, 1132 
\bibitem{} Patterson J.: 2011, MNRAS, 411,2695
\bibitem{} Pretorius M.~L., Knigge C.: 2012, MNRAS, 419, 1442
\bibitem{} Pretorius M.~L., Knigge C.: 2008a, MNRAS, 385, 1471 
\bibitem{} Pretorius M.~L., Knigge C.: 2008b, MNRAS, 385, 1485 
\bibitem{} Pretorius M.~L., Knigge C., Kolb U.: 2007a, MNRAS, 374, 1495 
\bibitem{} Pretorius, M.~L., Knigge, C., Schwope, A.~D.: 2013, MNRAS, 432, 570 
\bibitem{} Pretorius M.~L., et al.: 2007b, MNRAS, 382, 1279 
\bibitem{} Ritter H., Kolb U.: 2003, A\&A, 404, 301
\bibitem{} Schwope A.D., et al.: 2002, A\&A, 396, 895 
\bibitem{} Tout C.~A., et al.: 2008, MNRAS, 387, 897
\bibitem{}Townsley D.~M., G{\"a}nsicke B.~T.: 2009, ApJ, 693, 1007 
\bibitem{}Wickramasinghe D.~T., Wu K., Ferrario L.: 1991, MNRAS, 249, 460 
\bibitem{} Warner B.: 1987, MNRAS, 227, 23
\bibitem{} van Teeseling A., Beuermann K., Verbunt F.: 1996, A\&A, 315, 467 



\end{thebibliography}
\end{document}